\documentclass[12pt]{article}
\usepackage{authblk, cite, color}
\usepackage{amssymb, amsmath}
\usepackage[toc]{appendix}
\usepackage[colorlinks=true,linkcolor=red]{hyperref}
\usepackage[dvipsnames]{xcolor}
\begin{document}
\title{\bf Geometrical Applications of Split Octonions}
\author[1,2]{Merab Gogberashvili \thanks{gogber@gmail.com}}
\author[1]{Otari Sakhelashvili \thanks{otosaxel@gmail.com}}
\affil[1]{\small Tbilisi Ivane Javakhishvili State University, 3 Chavchavadze Ave., Tbilisi 0179, Georgia}
\affil[2]{\small Andronikashvili Institute of Physics, 6 Tamarashvili St., Tbilisi 0177, Georgia}
\maketitle

\begin{abstract}

It is shown that physical signals and space-time intervals modeled on split-octonion geometry naturally exhibit properties from conventional (3+1)-theory (e.g. number of dimensions, existence of maximal velocities, Heisenberg uncertainty, particle generations, etc.). This paper demonstrates these properties using an explicit representation of the automorphisms on split-octonions, the noncompact form of the exceptional Lie group G2. This group generates specific rotations of (3+4)-vector parts of split octonions with three extra time-like coordinates and in infinitesimal limit imitate standard Poincare transformations. In this picture translations are represented by non-compact Lorentz-type rotations towards the extra time-like coordinates. It is shown how the G2 algebra's chirality yields an intrinsic left-right asymmetry of a certain 3-vector (spin), as well as a parity violating effect on light emitted by a moving quantum system. Elementary particles are connected with the special elements of the algebra which nullify octonionic intervals. Then the zero-norm conditions lead to free particle Lagrangians, which allow virtual trajectories also and exhibit the appearance of spatial horizons governing by mass parameters.

\end{abstract}
\vskip 5mm


\section{Introduction}

Many properties of physical systems can be revealed from the analysis of proper mathematical structures used in descriptions of these systems. The geometry of space-time, one of the main physical characteristics of nature, can be understood as a reflection of symmetries of physical signals we receive and of the algebra used in the measurement process. Since all observable quantities we extract from single measurements are real, in geometrical applications it is possible to restrict ourselves to the field of real numbers. To have a transition from a manifold of the results of measurements to geometry, one must be able to introduce a distance between some objects and an etalon (unit element) for their comparison. In algebraic language all these physical requirements mean that to describe geometry we need a composition algebra with the unit element over the field of real numbers.

Besides usual real numbers, according to the Hurwitz theorem, there are three unique normed division algebras -- complex numbers, quaternions and octonions \cite{Sc, Sp-Ve, Baez}. Physical applications of the widest normed algebra of octonions are relatively rare (see reviews \cite{Rev-1, Rev-2, Rev-3, Rev-4, Rev-5}). One can point to the possible impact of octonions on: Color symmetry \cite{Gun-Gur, Color-1, Color-2}; GUTs \cite{GUT-1, GUT-2, GUT-3, GUT-4}; Representation of Clifford algebras \cite{Cliff-1, Cliff-2, Cliff-3}; Quantum mechanics \cite{QM-1, QM-2, QM-3, QM-4, QM-5, QM-6}; Space-time symmetries \cite{Rel-1, Rel-2}; Field theory \cite{QFT-1, QFT-2, QFT-3}; Quantum Hall effect \cite{Hall}; Kaluza-Klein program without extra dimensions \cite{KK-1, KK-2, KK-3}; Strings and M-theory \cite{String-1,String-2, String-3, String-4, String-5}; SUSY \cite{SUSY}; etc.

The essential feature of all normed composition algebras is the existence of a real unit element and a different number of hypercomplex units. The square of the unit element is always positive, while the squares of the hypercomplex basis units can be negative as well. In physical applications one mainly uses division algebras with Euclidean norms, whose hypercomplex basis elements have negative squares, similar to the ordinary complex unit $i$. The introduction of vector-like basis elements (with  positive squares) leads to the split algebras with pseudo-Euclidean norms.

In this paper we propose parameterizing world-lines (paths) of physical objects by the elements of real split octonions \cite{Gog},
\begin{equation} \label{s}
s = \omega + \lambda^nJ_n + x^nj_n + ct I~. ~~~~~ (n = 1, 2, 3)
\end{equation}
A pair of repeated upper and lower indices, by the standard convention, implies a summation, i.e. $x^nj_n = \delta_{nm}x^nj^m$, where $\delta^{nm}$ is Kronecker's delta.

In geometric application four of the eight real parameters in (\ref{s}), $t$ and $x^n$, are space-time coordinates, and $\omega$ and $\lambda^n$ are interpreted as the phase (classical action) and the wavelengths associated with the octonionic signals \cite{Gog}.

The eight basis units in (\ref{s}) are represented by one scalar (denoted by $1$), the three vector-like objects $J_n$, the three pseudovector-like elements $j_n$ and one pseudoscalar-like unit $I$. The squares (inner products) of seven of the hypercomplex basis elements of split octonions give the unit element, $1$, with the different signs,
\begin{equation} \label{JjI}
J_n^2=1~, ~~~~~ j_n^2=-1~, ~~~~~ I^2=1~.
\end{equation}
It is known that to generate a complete basis of split octonions the multiplication and distribution laws of only three vector-like elements $J_n$ are enough \cite{Sc, Sp-Ve, Baez}. The three pseudovector-like basis units, $j_n$, can be defined as the binary products,
\begin{equation} \label{jI}
j_n = \frac{1}{2} \varepsilon_{nmk}J^mJ^k ~~~~~(n,m,k = 1,2,3)
\end{equation}
($\varepsilon_{nmk}$ is the totally antisymmetric unit tensor of rank three), and thus describes orthogonal planes spanned by two vector-like elements $J_n$. The seventh basis unit $I$ (the oriented volume) is defined as the triple product of all three vector-like elements and has three equivalent representation in terms of $J^n$ and $j^n$,
\begin{equation} \label{I}
I = J_1j_1 = J_2j_2 = J_3j_3 ~.
\end{equation}
So the algebra of all non-commuting hypercomplex basis units of split octonions has the form:
\begin{eqnarray} \label{algebra}
J_nJ_m &=& - J_mJ_n = \varepsilon_{nmk} j^k,\nonumber\\
j_nj_m &=& -j_mj_n = \varepsilon_{nmk} j^k,\nonumber\\
j_mJ_n &=& - J_nj_m = \varepsilon_{nmk}J^k, \\
J_nI &=& - IJ_n = j_n~,\nonumber\\
j_nI &=& -Ij_n = J_n~.\nonumber
\end{eqnarray}

Conjugations of octonionic basis units, which can be understand as the reflection of vector-like elements,
\begin{equation} \label{bar-J}
J_n^\dag = - J_n ~,
\end{equation}
reverses the order of $J_n$ in products, i.e.
\begin{eqnarray} \label{bar}
j_n^\dag &=& \frac 12 \left(\varepsilon_{nmk}J^{m}J^{k}\right)^\dag = \frac 12 \varepsilon_{nmk}J^{k\dag}J^{m\dag} = - j_n~, \nonumber \\
I^\dag &=& \left(J_1 J_2 J_3\right)^\dag = J_3^\dag J_2^\dag J_1^\dag= - I~.
\end{eqnarray}
So the conjugation of (\ref{s}) gives
\begin{equation} \label{s*}
s^\dag = \omega - \lambda_n J^n - x_n j^n - ct I~.
\end{equation}

Using (\ref{JjI}), (\ref{algebra}) and (\ref{s*}) one can find that the norm of (\ref{s}) is given by,
\begin{equation} \label{sN}
N^2 = ss^\dag = s^\dag s = \omega^2 - \lambda^2 + x^2 - c^2t^2
\end{equation}
(where $\lambda^2 = \lambda_n\lambda^n$ and $x^2 = x_nx^n$), represents some kind of 8-dimensional space-time with (4+4)-signature and reduces to the classical formula of Minkowski intervals if $\omega^2 - \lambda^2 = 0$.

As for the case of ordinary Minkowski space-time, we assume that for physical events the corresponding 'intervals' given by (\ref{sN}) are non-negative. A second condition is that for physical signals the vector part of split octonions (\ref{s}) should be time-like,
\begin{equation} \label{time-like}
c^2t^2 + \lambda_n\lambda^n > x_nx^n~.
\end{equation}
The classification of split octonions by the values of their norms is presented in \ref{Classification}.


\section{Transformations and automorphisms}

To find the geometry associated with the signals (\ref{s}) we need to represent rotations by split octonions, since due to non-associativity of the algebra the group of rotations of octonionic axes, $1$, $J_n$, $j_n$, and $I$ ($G_2^{NC}$, is not equivalent to the group of accompanied passive tensorial transformations of coordinates, $\omega$, $\lambda_n$, $x_n$ and $t$ ($SO(4,4)$). In \ref{One-side} it is shown that by rotations in four orthogonal planes of octonionic space (one of which always includes the axis of real numbers $\omega$) the octonion (\ref{s}) always can be represented as the sum of four elements. The left multiplication, $Rs$, by a octonion with unit norm (\ref{Classification}),
\begin{equation} \label{R}
R = e^{\epsilon \theta}~,
\end{equation}
where $\epsilon$ is the (3+4)-vector defined by seven basis units $J_n$, $j_n$ and $I$, represents rotations by the angles $\theta$ in these four planes. The right product, $sR^{-1}$, will reverse the direction of rotation in the plane of $\omega$. So we can represent a rotation, which only affects the (3+4)-vector part of $s$, applying the unit half-angle octonion twice, multiplying on both the left and on the right (with its inverse),
\begin{equation} \label{s'}
s' = R s R^{-1} = R s R^\dag = e^{\epsilon\theta/2 } s e^{-\epsilon\theta/2}~.
\end{equation}
These maps are well-defined, since the associator of the triplet $(R,s,R^{-1})$ vanishes. The set of rotations (\ref{s'}) of the seven octonionic coordinates $x^n, \lambda^n$ and $t$, in three planes, which do not affect the scalar part $\omega$ of (\ref{s}), contains $7 \times 3 = 21$ angles of the group $SO(3,4)$ of passive transformations of coordinates.

To represent the active rotations in the space of $s$, which preserve the norm (\ref{sN}) and multiplicative structure of octonions (\ref{algebra}) as well, we would need the transformations to be automorphisms (\ref{Automorphisms}). An automorphism $U$ of any two octonions, $O_1$ and $O_2$, gives the equation:
\begin{equation} \label{U}
\left( U O_1 U^{-1}\right) \left( U O_2 U^{-1}\right) =  U O_1O_2 U^{-1} ~,
\end{equation}
which only holds in general if multiplication of $U$, $O_1$ and $O_2$ is associative. Combinations of the rotations (\ref{s'}) around different octonionic axes are not unique. This means that not all transformations of $SO(3,4)$ form a group and can be considered as real rotations. Only the transformations that have a realization as associative multiplications should be considered. Also note that an automorphism of split octonions can be generated only by the octonions with positive norms (\ref{Classification}), since the set of split octonions with negative norms do not form a group (it is not closed under multiplication). So to model space-time symmetries we need to use the group of automorphisms of split octonions with positive norms.

It is known that associative transformations of split octonions can be done by the specific simultaneously rotations in two (and not in three as for $SO(3,4)$) orthogonal octonionic planes (\ref{Automorphisms}). These rotations form a subgroup of $SO(3,4)$ with $2\times 7 = 14$ parameters, known as the automorphism group of split octonions, $G_2^{NC}$. Generators of this real non-compact form of Cartan's smallest exceptional Lie group $G_2$ is presented in \ref{Generators}.

Infinitesimal transformations of coordinates which accompany the group of active transformations of octonionic basis units, $G_2^{NC}$, and preserve the diagonal quadratic form, $x^2 - \lambda^2 - dt^2 \geq 0$, can be written in the form (\ref{Automorphisms}):
\begin{eqnarray} \label{x-nu}
x_n' &=& x_n - \varepsilon_{nmk} \alpha^m x^k - \theta_n ct + \frac 12 \left( |\varepsilon_{nmk}|\phi^m + \varepsilon_{nmk} \theta^m \right) \lambda^k + \left(\varphi_n - \frac 13 \sum_m \varphi_m\right) \lambda_n~, \nonumber \\
ct' &=& ct - \beta_n \lambda ^n - \theta_nx^n ~, \\
\lambda_n' &=& \lambda_n - \varepsilon_{nmk} \left(\alpha^m - \beta^m\right) \lambda^k + \beta_n ct + \frac 12 \left( |\varepsilon_{nmk}| \phi^m - \varepsilon_{nmk} \theta^m \right) x^k + \left(\varphi_n - \frac 13 \sum_m \varphi_m\right) x_n~, \nonumber
\end{eqnarray}
with no summing over $n$ in the last terms of $x_n'$ and $\lambda_n'$. From the five 3-angles in (\ref{x-nu}), $\alpha^m$, $\beta^m$, $\phi^m$, $\theta^m$ and $\varphi^m$, only 14 are independent because of the condition:
\begin{equation}
\sum_n \left(\varphi_n - \frac 13 \sum_m \varphi_m\right) = 0~.
\end{equation}

The Lorentz-type transformations (\ref{x-nu}) of the 7-dimensional space with four time-like coordinates should describe space-time symmetries if the split octonions form relevant algebraic structures for microphysics. The transformations (\ref{x-nu}) formally can be divided in three distinct classes (\ref{Automorphisms}): Euclidean rotations of the spatial, $x_n$, and time-like ($t$ and $\lambda_n$) coordinates by the compact 3-angles $\alpha^n$ and $\beta^n$, respectively; Boosts, mixing of spatial and time-like coordinates by the two hyperbolic 3-angles  $\theta^n$ and $\phi^n$; Diagonal boosts of the spatial coordinates, $x_n$, and corresponding time-like parameters, $\lambda_n$, by the hyperbolic angles $\varphi^n$.

We notice that if we consider rotations by the angles $\alpha^n$, $\beta^n$ and $\theta^n$, i.e. assume that
\begin{equation}
\phi^m = \varphi^m = 0~,
\end{equation}
the passive $G_2^{NC}$-transformations (\ref{x-nu}) of only ordinary space-time coordinates, $x_n$ and $t$, will imitate the ordinary infinitesimal Poincar\'{e} transformations of (3+1)-Minkowski space,
\begin{equation} \label{Lorentz}
x_n' = x_n - \varepsilon_{nmk} \alpha^m x^k - \theta_n ct + a_n~, ~~~~~ ct' = ct  - \theta_nx^n + a_0 ~.
\end{equation}
Here the space-time translations:
\begin{equation}
a_n = \frac 12  \varepsilon_{nmk} \theta^m  \lambda^k~, ~~~~~ a_0 = - \beta_n \lambda ^n~,
\end{equation}
are generated by the Lorentz-type rotations toward the time-like directions $\lambda_n$. So in the language of octonionic geometry any motion in ordinary space-time is generated by $\lambda^n \sim p^n/p^2$. Time translations $a_0$ are smooth, since $\beta_n$ are compact angles. However, the angles $\theta^m$ are hyperbolic and for any active spatial translation $a_n$ there exist a horizon (analogues to the Rindler horizon), what is equivalent to the introduction of some mass scale, or inertia (see Sect. 6).

For completeness note that there exists a second well-known representation of $G_2^{NC}$ as the symmetry group of a ball rolling on a larger fixed ball without slipping or twisting, when the ratio of the balls radii is $1/3$ \cite{Ball-1, Ball-2, Ball-3, Ball-4, Ball-5}. Understanding the exceptional Lie groups as the symmetry groups of naturally occurring objects is a long-standing program in mathematics. Symmetries of rolling balls are visualized in (\ref{x-nu}) by the presence of two 3-vectors, $x_n$ and $\lambda_n$, one of which is time-like. The fourth time-like coordinate (ordinary time $t$), which also is affected by (\ref{x-nu}), breaks the symmetry between 'balls' and the factor $1/3$ of the ratio of their radii corresponds to the existence of the three extra time-like coordinates $\lambda_n$.


\section{Boosts, Heisenberg uncertainty and chirality}

In the algebra of split octonions there exist only three space-like parameters, $x_n$, and three independent compact rotations around corresponding axes, what explains why physical space described by octonionic signals has three spatial dimensions. Let us analyze new features of the $G_2^{NC}$-transformations (\ref{x-nu}) in comparison with standard Lorentz's formulas for (3+1)-Minkowski space.

Euclidean rotations around one of the space-like axes, $x_1$, the automorphism (\ref{rotation}), correspond to the following passive infinitesimal transformations of coordinates:
\begin{eqnarray} \label{al-be}
&x_1' = x_1~, ~~~~~ x_2' = x_2 + \alpha_1 x_3~, ~~~~~ x_3' = x_3 - \alpha_1 x_2~, \nonumber \\
&ct' = ct - \beta_1 \lambda_1~, \\
&\lambda_1' = \lambda_1 + \beta_1ct~, ~~~~~ \lambda_2' = \lambda_2 + (\alpha_1 - \beta_1) \lambda_3~, ~~~~~\lambda_3' = \lambda_3 - (\alpha_1 - \beta_1) \lambda_2~.\nonumber
\end{eqnarray}
When $\alpha_1 = \beta_1$ this reduces to the standard rotation by the Euler angle $\alpha_1$, which also causes translations of time, $t$, due to mixings with the extra time-like parameter $\lambda_1$. In general, any active Euclidean 3-rotation of the spatial coordinates $x^n$ changes the time parameter by the amount of the corresponding $\lambda_n$, which can be understood as passing of time in our world.

Analysis of boosts of $G_2^{NC}$ can help us in the physical interpretation of the extra time-like parameters $\lambda_n$. Consider the automorphisms (\ref{boost}) by the hyperbolic angles $\theta_1$ and $\phi_1$,
\begin{eqnarray} \label{theta-1}
&x_1' = x_1 - \theta_1 ct~, ~~~~~ x_2' = x_2 + \frac 12 \left(\phi_1 - \theta_1\right) \lambda_3~, ~~~~~ x_3' = x_3 + \frac 12 \left(\phi_1 + \theta_1\right)  \lambda_2~, \nonumber \\
&ct' = ct - \theta_1 x_1~, \\
&\lambda_1' = \lambda_1 ~, ~~~~~ \lambda_2' = \lambda_2 + \frac 12 \left(\phi_1 + \theta_1\right)  x_3~, ~~~~~\lambda_3' = \lambda_3 + \frac 12 \left(\phi_1 - \theta_1\right)  x_2~. \nonumber
\end{eqnarray}
The case $\phi_1 = \theta_1$ corresponds to ordinary boost in $(t,x)$-planes -- transitions to the reference frame moving with the velocity $\theta_1$ along the axis $x_1$. Then if we consider the motion of the origin of the moving system,
\begin{equation} \label{x=y=z=0}
x' = y' = z' = 0 ~,
\end{equation}
from the first line in (\ref{theta-1}) we find that
\begin{equation}
\theta_1 = \frac {x_1}{ct} ~, ~~~~~ \lambda_3 = \frac {x_2}{\theta_1} ~, ~~~~~ \lambda_2 = - \frac {x_3}{\theta_1} ~,
\end{equation}
i.e. the quantities $\lambda_2$ and $\lambda_3$ are inversely proportional to the velocity, $x_1/t$. In the space of split octonions (\ref{s}) there are two classes of time-like parameters ($t$ and $\lambda_n$) and two different light-cones. So, in addition to $c$, there must exist the second fundamental constant, which can be extracted from $\lambda_n$. In quantum mechanics the quantity with the dimension of length, which is proportional to a fundamental physical constant and is inversely proportional to velocity (or momentum, $p_n$), is called the wavelength. So it is natural to assume that
\begin{equation} \label{l=1/p}
\lambda^n = \hbar \frac {p^n}{p^2} \sim \frac {\hbar}{p_n}~,
\end{equation}
where $\hbar$ is the Planck constant and $p^n$ is the 3-momentum associated with the octonionic signal. So in our approach the two fundamental physical constants, $c$ and $\hbar$, have geometrical origin and correspond to two kinds of light-cone signals in the space of split octonions (\ref{sN}).

When $\phi_1 = 0$ from (\ref{theta-1}) we find that the ratios,
\begin{equation} \label{DeltaX/l}
\frac{\Delta x_2}{\Delta \lambda_2} = - \frac {\lambda_3}{x_3}~, ~~~~~ \frac{\Delta x_3}{\Delta \lambda_3} = - \frac {\lambda_2}{x_2}
\end{equation}
(where $\Delta x_n = x'_n - x_n$ and $\Delta \lambda_n = \lambda'_n - \lambda_n$), are unchanged under infinitesimal transformations (\ref{theta-1}). Similar relations can be obtained for the boosts along two other space-like axes. From the invariance of octonionic norms (\ref{s-I}) we know that
\begin{equation} \label{l/x>1}
|x_n| \gtrsim |\lambda_n|~.
\end{equation}
Then inserting (\ref{l=1/p}) and (\ref{l/x>1}) into (\ref{DeltaX/l}) we can conclude that the uncertainty relations,
\begin{equation} \label{Un}
\Delta x_n \Delta p_n \geq \hbar ~,
\end{equation}
in our model has the same geometrical meaning as the existence of the maximal velocity, $c$ \cite{Gog, Gog-Split}.

From (\ref{theta-1}) we also notice that in the planes orthogonal to $v_1= x_1/t$ the pair $x_2'$, $\lambda_3'$ increases, while $x_3'$, $\lambda_2'$ decreases, and for the case (\ref{x=y=z=0}) we have,
\begin{equation} \label{x/l}
\frac {x_2}{\lambda_3} = - \frac {x_3}{\lambda_2} = \theta_1~.
\end{equation}
Using similar relations for the boosts along other two spatial directions we conclude that there exists some 3-vector (spin),
\begin{equation}
\sigma_n = \frac {1}{\hbar} \varepsilon_{nmk} x^m p^k \sim \varepsilon_{nmk}\frac {x^m}{\lambda_k}~,
\end{equation}
which characterizes relative rotations of $x_n$ and $\lambda_n$ in the direction orthogonal to the velocity planes. One can speculate on the connection of the left-right asymmetry in relative rotations of $x_n$ and $\lambda_n$ in (\ref{theta-1}) with the right-handed neutrino problem for massless case.


\section{Parity violation}

For the reference frame (\ref{x=y=z=0}), for the finite angles $\phi_1 = \theta_1$, from (\ref{theta-1}) one can obtain the standard relativistic expressions:
\begin{equation}
\tanh \theta_1 = \frac {\cal V}c~, ~~~~~ \cosh \theta_1 = \frac {1}{\sqrt{1 - {\cal V}^2/c^2}}~, ~~~~~ \sinh \theta_1 = \frac {{\cal V}/c}{\sqrt{1 - {\cal V}^2/c^2}}~,
\end{equation}
where ${\cal V}$ is the velocity of moving system along $x_1$. Then from (\ref{theta-1}) there follows the generalized rule of velocity addition,
\begin{equation} \label{v'}
v_1'=\frac{v_1 - c\tanh \theta_1}{1 - \tanh \theta_1 \frac {v_1}{c}}~, ~~~~~
v_2'=\frac{v_2  - \tanh \theta_1 \dot\lambda_3}{1 - \tanh \theta_1 \frac {v_1}{c}}~, ~~~~~
v_3'=\frac{v_3 + \tanh \theta_1 \dot\lambda_2}{1 - \tanh \theta_1 \frac {v_1}{c}}~.
\end{equation}
We see that the standard expressions are altered only in the planes orthogonal to ${\cal V}$, and that what is important are the terms with different sign. One consequence of this fact is that the formula for the aberration of light will be modified.

Consider photons moving in the $(x_1,x_2)$-plane,
\begin{eqnarray}
v_1 = c \cos \gamma_{12}~, ~~~~~ v_2 = c \sin \gamma_{12} ~, \nonumber \\
v'_1 = c \cos \gamma_{12}'~, ~~~~~ v'_2 = c \sin \gamma_{12}' ~,
\end{eqnarray}
where $\gamma_{12}$ and $\gamma_{12}'$ are angles between $v_1$ and $v'_1$ and the axes $x_1$ and $x'_1$, respectively. Using (\ref{v'}), for the case ${\cal V}/c \ll 1$, we have,
\begin{equation}
\sin \gamma_{12}' = \frac {\sin \gamma_{12} - \frac {\cal V}{c^2} \dot \lambda_3}{1 - \frac {\cal V}c \cos \gamma_{12}}~, ~~~~~ \cos \gamma_{12}' = \frac {\cos \gamma_{12} - \frac {\cal V}{c}}{1 - \frac {\cal V}c \cos \gamma_{12}}~,
\end{equation}
where $\dot \lambda_3$ is the rate of Doppler's shift along an axis,$x_3$, orthogonal to the photon's trajectory. Then angle of aberration in the $(x_1,x_2)$-plane give the value:
\begin{equation} \label{gamma-12}
\Delta \gamma_{12} = \gamma_{12}' - \gamma_{12} = \frac{\cal V}c \sin \gamma_{12} - \frac {\cal V}{c^2} \dot \lambda_3~.
\end{equation}
Analogous to (\ref{gamma-12}) the formula for the aberration angle in the $(x_1,x_3)$-plane has the form:
\begin{equation} \label{gamma-13}
\Delta \gamma_{13} = \frac {\cal V}c \sin \gamma_{13} + \frac {\cal V}{c^2} \dot \lambda_2~,
\end{equation}
where $\dot \lambda_2$ is the rate of the Doppler shift towards the orthogonal to the $(x_1,x_3)$-plane direction. This spatial asymmetry of aberration, which distinguishes the left and right coordinate systems, may be detectable by precise quantum measurements.


\section{Spin and hypercharge}

Now consider the last class of automorphisms (\ref{Automorphisms}),
\begin{equation} \label{varphi}
x_n' = x_n + \left(\varphi_n - \frac 13 \sum_m \varphi_m \right) \lambda_n~, ~~~~~ t' = t ~, ~~~~~ \lambda_n' = \lambda_n  + \left(\varphi_n - \frac 13 \sum_m \varphi_m \right) x_n~,
\end{equation}
with no summing by $n$. These transformations represent rotations of the three pairs of space-like and time-like coordinates, $(x_1, \lambda_1)$, $(x_2, \lambda_2)$ and $(x_3,\lambda_3)$, into each other. We have the three planes, $(x_1 - \lambda_1)$, $(x_2 - \lambda_2)$ and $(x_3 - \lambda_3)$, that undergo rotations through hyperbolic angles, $\varphi_1$, $\varphi_2$ and $\varphi_3$ (of the three only two are independent), which are the only hyper-planes in the space of split octonions that are affected by one and not two 3-angles of automorphisms. So it is natural to define the Abelian subalgebra of $G_2^{NC}$ by generators of two independent rotations in these planes. It is known that the rank of the Cartan's subalgebra of $G_2^{NC}$ is the same as of the group $SU(3)$ \cite{BHW, Gun-Gur}. Indeed, in terms of two parameters, $K_1$ and $K_2$, which are related to the angles $\varphi_n$ and (\ref{k-varphi}) as:
\begin{equation} \label{K}
K_1 = k_1 = \frac 13 \left(2\varphi_1 - \varphi_2 - \varphi_3\right)~, ~~~~~ K_2 = \frac {\sqrt 3}{2}\left(k_1 + k_1\right) = -  \frac {1}{2\sqrt 3} \left(2 \varphi_3 - \varphi_1 - \varphi_2 \right)~,
\end{equation}
the transformations (\ref{varphi}) can be written more concisely,
\begin{equation} \label{l-x}
\begin{pmatrix}
\lambda_1' + I x_1' \cr
\lambda_2' + I x_2' \cr
\lambda_2' + I x_2'
\end{pmatrix}
= e^{\left(K_1\Lambda_3 + K_2 \Lambda_8\right) I}
\begin{pmatrix}
\lambda_1 + I x_1 \cr
\lambda_2 + I x_2 \cr
\lambda_2 + I x_2
\end{pmatrix}~,
\end{equation}
where $I$ is the vector-like octonionic basis unit ($I^2 = 1$) and $\Lambda_3$ and $\Lambda_8$ are the standard $3\times 3$ Gell-Mann matrices,
\begin{equation}
\Lambda_3 =
\begin{pmatrix}
1 & 0 & 0 \cr
0 & -1 & 0\cr
0 & 0 & 0
\end{pmatrix}
~, ~~~~~ \Lambda_8 = \frac {1}{\sqrt 3}
\begin{pmatrix}
1 & 0 & 0 \cr
0 & 1 & 0\cr
0 & 0 & -2
\end{pmatrix}~.
\end{equation}
Then, using analogies with $SU(3)$, one can classify irreducible representations of the space-time group of our model, $G_2^{NC}$, by the two fundamental simple roots ($K_1$ and $K_2$) corresponding to spin and hypercharge. It is known that all quarks, antiquarks, and mesons can be imbedded in the adjoint representation of $G_2^{NC}$ \cite{Gun-Gur}. So the symmetry (\ref{varphi}), in addition to the uncertainty relations, probably shows the existence of three generations of objects which are necessary to extract three spatial coordinates from any octonionic signal $s$. To clarify this point note that (\ref{sN}) can be viewed as some kind of space-time interval with four time-like dimensions. The ordinary time parameter, $t$, corresponds to the distinguished octonionic basis unit, $I$, while the other three time-like parameters, $\lambda_n$, have a natural interpretation as wavelengths, i.e. do not relate to time as conventionally understood. It is known that a unique physical system in multi-dimensional geometry generates a large variety of 'shadows' in (3+1)-subspace as different dynamical systems (in terms of different Hamiltonians) \cite{times-1, times-2, times-3, times-4, times-5, times-6, times-7}. The information of multi-dimensional structures, which is retained by these images of the initial system, takes the form of hidden symmetries. For the case of fundamental physical signals with three extra time-like dimensions, in addition to massless particles (which are not affected by extra times), one can observe three generations of particles with different mass (corresponding to rotations of $t$ in one, two, or all three extra time-like planes), see (\ref{sDec}).  Note the factor $1/3$ in front of $\omega$ (action) in the first equation (\ref{sDec}) that appears due to existence of the three extra time-like parameters, $\lambda_n$. Then three independent $(\omega, \lambda_n)$-rotations in 8-dimensional octonionic space will give the appearance of three generations of particles in ordinary (3+1)-dimensions. Such a possibility is not ruled out by problems with ghosts and with unitarity, since split octonions with positive norms form the division algebra.


\section{Free particle Lagrangians}

It is known that in split algebras there can be constructed special elements with zero norms, which are called zero divisors \cite{Sc}. This zero norm objects are important structures in physical applications \cite{So}. Zero divisors (light-cone operators) of split octonions could serve as the unit signals and thus may describe elementary particles. Then the number of primitive idempotents (eight) and nilpotents (twelve) numerate the types of fundamental particles (\ref{Zero}), bosons and fermions, respectively. Indeed, the properties that the product of two projection operators reduces to the same idempotent (\ref{DD}), while the product of two Grassmann numbers is zero (\ref{GG}), naturally explains the validity of Bose and Fermi statistics for the corresponding elementary particles.

The zero-norm condition, $N^2 = 0$, for the paths (\ref{sN}) of elementary particles, be they massless or massive, can be written in the form:
\begin{equation} \label{ds}
ds^2 = ds ds^\dag = d\omega^2 - c^2 dt^2 \left(1 - \frac{v^2}{c^2} + \frac{\dot\lambda^2}{c^2}\right)= 0~.
\end{equation}

For photons, $|v| = c$, the condition (\ref{ds}) is equivalent to the Eikonal equation:
\begin{equation} \label{Eikonal}
\frac{d\omega}{d\lambda_n}\frac{d\omega}{d\lambda^n} = 1~,
\end{equation}
where the wavelengths, $\lambda_n \sim \hbar / p_n$, serve as the parameters of trajectory. The relation (\ref{Eikonal}) justifies the interpretation of $\omega$ as the phase function (frequency) corresponding to the classical action of particles.

For massive particles the time coordinate, $t$, is good parameter to describe motion and from (\ref{ds}) we find that the scalar part of a split octonionic signal, $\omega$, which is unchanged under automorphisms of $G_2^{NC}$, is the conserved function:
\begin{equation} \label{Action}
\frac{d\omega}{d t} = 0~.
\end{equation}
Then (\ref{ds}) can be written as:
\begin{equation} \label{omega=A}
\omega = \frac {A}{mc} = -c\int dt \sqrt{1 - \frac{v^2}{c^2} + \frac{\dot\lambda^2}{c^2}}~,
\end{equation}
where $A$ is the classical action of the particle with the mass $m$. So using (\ref{l=1/p}) the one-particle Lagrangian,
\begin{eqnarray} \label{L}
L = - mc^2\sqrt{1 - \frac{v^2}{c^2} + \hbar^2 \frac{\dot{p}^2}{c^2p^4}}~,
\end{eqnarray}
contains an extra 'quantum' term, which may be relevant for the relativistic velocities, $|v| \sim c$, or for the case of large forces,
\begin{equation}
\dot{p}_n \sim \frac {cp^2}{\hbar}~.
\end{equation}
From (\ref{L}) we see that on small distance scales, $\sim \hbar/|p|$, a particle can even exceed the speed of light (become virtual), since the new quantum term in (\ref{L}) is positive. The condition (\ref{ds}) and the invariance of the classical action (\ref{omega=A}) under $G_2^{NC}$ hold only for real trajectories. However, there exist the larger group of invariance of the interval (\ref{sN}), $SO(4,4)$, the group of passive tensorial transformations of all eight octonionic coordinates, which in general mixes space-like and time-like subspaces and thus introduces virtual trajectories of particles.

Applying the condition (\ref{time-like}) (that for physical signals the vector part of a split octonion should be time-like) to (\ref{L}) its follows that there should exists some maximal force:
\begin{equation} \label{dot-p}
\dot{|p|} \le |v|\frac {p^2}{\hbar} = m^2 \frac {c^3}{\hbar}~,
\end{equation}
where $m$ denotes the maximal possible value for the mass of particle associated with physical signals. Using estimations of  \cite{Gibb, Schi} for the maximum force, $F_{max} = c^4/4G$, where $G$ is the gravitational constant, from (\ref{dot-p}) we find that the maximal mass of particles associated with any octonionic signal is the Planck mass:
\begin{equation}
m^2 \sim \frac{\hbar c}{G}~.
\end{equation}


\section{Conclusion}

To conclude, in this paper it was analyzed consequences of describing physical signals in terms of split octonions over the field of real numbers. Eight real parameters of split octonions were related to space-time coordinates, the phase function (classical action) and wavelengths characterizing physical signals. The (4+4)-norms of the real split octonions are invariant under the group $SO(4,4)$ of tensorial transformation of parameters (passive transformations of coordinates), while the representation of rotations by octonions themselves (i.e. the active transformations of octonionic basis units) correspond to the real non-compact form of Cartan's exceptional Lie group $G_2^{NC}$ (the subgroup of $SO(4,4)$), which under certain conditions imitate the Poincar\'{e} transformations in ordinary space-time. Within this picture, in front of time-like coordinates in the expression of pseudo-Euclidean octonionic intervals there naturally appear two fundamental physical parameters, the light speed and Planck's constant. From the requirement of positive definiteness of norms under $G_2^{NC}$-transformations, together with the introduction of the maximal velocity, $c$, there follow conditions which are equivalent to uncertainty relations in quantum physics. By examining of $G_2^{NC}$-automorphisms it is explicitly shown that there exists of a 3-vector ("spin") that is fully chiral, which corresponds to the observed chirality of neutrinos in nature. We examined also the $G_2^{NC}$-effects on a quantum system moving at constant speed with respect to the observer, which emits light in various planes relative to the direction of motion, and new parity-violating effect on aberration and Doppler shift was derived. It was show how a particular class of $G_2^{NC}$-automorphisms of space-time coordinates can be expressed in concise form using two simple roots of the $SU(3)$ Lie algebra. Due to their conventional association with spin and hypercharge in physics, we argued that our split-octonion geometry inherently modeled three generations of fundamental particles. In our approach elementary particles are interpreted as light-cone operators, i.e. connected with the special elements of the algebra which nullify octonionic intervals. Then the zero-norm conditions in $G_2^{NC}$-geometry lead to corresponding free particle Lagrangians, which allow virtual particle trajectories (exceeding the speed of light, in certain cases) and exhibit the existence of spatial horizons governing by mass parameters.


\section*{Acknowledgments:}

This research was supported by the Shota Rustaveli National Science Foundation grant ${\rm ST}09\_798\_4-100$. O.S. acknowledges also the scholarship of World Federation of Scientists.


\appendix
\gdef\thesection{Appendix \Alph{section}}
\renewcommand{\theequation}{\Alph{section}.\arabic{equation}}


\section{Classification of split octonions} \label{Classification}
\setcounter{equation}{0}

The algebra of split octonions is a non-commutative, non-associative, non-division ring. Any element (\ref{s}) of the ring can be represented as:
\begin{equation} \label{S+V}
s = \omega + V~, ~~~~~ s^\dag = \omega - V~,
\end{equation}
where the symbols $\omega$ and
\begin{equation} \label{V}
V = \lambda^nJ_n + x^nj_n + ct I ~~~~~ (n = 1, 2, 3)
\end{equation}
are called the scalar and vector parts of octonion, respectively. Similar to the case of split quaternions \cite{Gog-Split, Oz-Er}, there exist three classes of split octonions having positive, negative or zero norm,
\begin{equation} \label{N=o-V}
N^2 = ss^\dag = s^\dag s = \omega^2 - V^2~,
\end{equation}
where
\begin{equation} \label{V2}
V^2 = - x^2 + c^2t^2 + \lambda^2 ~.
\end{equation}
In addition one can distinguish split octonions with time-like, space-like and light-like vector parts (\ref{V}):
\begin{eqnarray}
V^2 &<& 0 ~, ~~~~~ (space-like) \nonumber \\
V^2 &>& 0 ~, ~~~~~ (time-like) \\
V^2 &=& 0 ~. ~~~~~ (light-like)\nonumber
\end{eqnarray}

Split octonions with non-zero norms, $\omega^2 \neq V^2$, have multiplicative inverses,
\begin{equation}
s^{-1} = \frac {s^\dag}{N^2}~,
\end{equation}
with the property:
\begin{equation}
ss^{-1}= s^{-1} s = 1~,
\end{equation}
while octonions with zero norm, $\omega^2 = V^2$, have no inverses.

One can construct also the polar form of a split octonion with non-zero norm,
\begin{equation}
s = N e^{\epsilon \theta} ~,
\end{equation}
where $\epsilon$ is a unit 7-vector ($\epsilon^2 = \pm 1$). For $N \neq 0$ the quantity
\begin{equation} \label{R=s/N}
R = \frac {s}{N}
\end{equation}
is called unit octonion, which is useful to represent rotations of (3+4)-vector spaces (\ref{V}).

Depending on the values of (\ref{N=o-V}) and (\ref{V2}) split octonions have three different representations:
\begin{itemize}
\item {Every split octonion with negative norm, $\omega^2 < V^2$, can be written in the form:
\begin{equation} \label{x-like}
s = N(\sinh \theta + \epsilon \cosh \theta)~,
\end{equation}
where
\begin{equation}
\sinh \theta = \frac {\omega}{N}~, ~~~~~ \cosh \theta = \frac {|V|}{N}~, ~~~~~\epsilon = \frac {\lambda^n J_n + x^n j_n + ct I} {|V|}~,
\end{equation}
and $\epsilon$ is a unit time-like 7-vector.}
\item {Every split octonion with positive norm, $\omega^2 > V^2$, and with time-like vector part, $V^2 > 0$, can be written in the form:
\begin{equation} \label{t-like-x}
s = N(\cosh \theta + \epsilon \sinh \theta)~,
\end{equation}
where
\begin{equation}
\cosh \theta = \frac {\omega}{N}~, ~~~~~ \sinh \theta = \frac {|V|}{N}~, ~~~~~\epsilon = \frac {\lambda^n J_n + x^n j_n + ct I}{|V|}~,
\end{equation}
so $\epsilon$ again is a unit time-like 7-vector.}
\item {Every split octonion with positive norm, $\omega^2 > V^2$, and with space-like vector part,  $V^2 < 0$, can be written in the form:
\begin{equation} \label{t-like-t}
s = N(\cos \theta + \epsilon \sin \theta)~,
\end{equation}
where
\begin{equation}
\cos \theta = \frac {\omega}{N}~, ~~~~~ \sin \theta = \frac {|V|}{N}~, ~~~~~\epsilon = \frac {\lambda^n J_n + x^n j_n + ct I}{|V|}~,
\end{equation}
thus $\epsilon$ now is a unit space-like 7-vector.}
\end{itemize}

The norm of a split octonion (\ref{N=o-V}) contains two types of 'light-cones', $N=0$ and $V=0$, and we need two constants that characterize the maximum spread angles of these cones. The first fundamental parameter, the speed of light $c$, appears in standard Lorentz boosts for the time-like signals, $V^2 > 0$. The second fundamental constant, $\hbar$, corresponds to maximal rotations in $(\omega, V)$-planes, when physical events have $N^2 \geq 0$ and still can be described by the actions $\sim \omega$.


\section{One-side products} \label{One-side}
\setcounter{equation}{0}

The 8-dimensional octonionic space (\ref{s}) always can be represented as the sum of four elements by rotations in four orthogonal planes (see, (\ref{s-j1}), (\ref{s-J1}) and (\ref{s-I}) below), one of which always includes the axis of real numbers $\omega$ \cite{Man-Sch}. Then it can be shown that multiplication of one octonion by another octonion of unit norm (\ref{R=s/N}) represents some kind of rotation in these four planes. In the algebra of split octonions there exist seven basis units, $J_n$, $j_n$ and $I$, which define the 7-vector $\epsilon$ in the formulas for one-sided (for example, the left) products,
\begin{equation} \label{Rs}
Rs = e^{\epsilon\theta/2 } s ~.
\end{equation}
So totally we have $4 \times 7 = 28$ parameters of the group $SO(4,4)$, which preserve the norm (\ref{sN}) \cite{Man-Sch}. Octonionic hypercomplex basis units have different algebraic properties and lead to the three distinct classes of the rotation operator $R$:
\begin{itemize}
\item {The three pseudovector-like basis elements, $j_n$, behave as the ordinary complex unit ($j_n^2 = -1$) and rotations around the corresponding spatial axes, $x_n$, can be done by the unit octonions with $N^2>0$ having space-like vector part:
\begin{equation} \label{R(j)}
e^{j_n\alpha_n } = \cos \alpha_n + j_n \sin \alpha_n ~,
\end{equation}
where $\alpha_n$ are three real compact angles. To show this explicitly let us consider the rotation operator $R$ in (\ref{Rs}) when $\epsilon = j_1$. In the octonionic algebra (\ref{algebra}) the basis unit $j_1$ has three different representations by other basis elements that form an associative triplets with $j_1$,
\begin{equation} \label{j1}
j_1 = J_2J_3 = j_2j_3 = J_1I~,
\end{equation}
or equivalently,
\begin{equation} \label{j_1}
j_1J_3 = J_2~, ~~~~~ j_1j_2 = j_3~, ~~~~~ j_1I = J_1~.
\end{equation}
Then it is possible to 'rotate out' four octonionic axes and rewrite (\ref{s}) in the equivalent form:
\begin{equation} \label{s-j1}
s = \sqrt{\omega^2 + x_1^2} e^{j_1 \alpha_\omega} + \sqrt{\lambda_2^2 + \lambda_3^2} e^{j_1 \alpha_\lambda}J_3 +\sqrt{x_2^2 + x_3^2} e^{j_1 \alpha_x}j_2 + \sqrt{t^2 + \lambda_1^2} e^{j_1 \alpha_t}I ~,
\end{equation}
where four angles are given by:
\begin{eqnarray}
&\cos \alpha_\omega = \frac {\omega}{\sqrt{\omega^2 + x_1^2}}~, ~~~~~ \cos \alpha_\lambda = \frac {\lambda_3}{\sqrt{\lambda_2^2 + \lambda_3^2}}~, \nonumber \\
&\cos \alpha_x = \frac {x_2}{\sqrt{x_2^2 + x_3^2}}~, ~~~~~ \cos \alpha_t = \frac {ct}{\sqrt{t^2 + \lambda_1^2}} ~.
\end{eqnarray}
Now it is obvious that when $R = e^{j_1\alpha_1}$ and $s$ is represented by (\ref{s-j1}), the product (\ref{Rs}) leads to simultaneous rotations in the four planes, $(\omega, x_1)$, $(\lambda_2, \lambda_3)$, $(x_2, x_3)$, $(\lambda_1, t)$, by the same compact angle $\alpha_1$. Similar results can be obtained for the rotations by the other two pseudo-vector like units, $j_2$ and $j_3$.}

\item {For the three vector-like basis elements, $J_n$, with the positive squares, $J_n^2 = 1$, we need the unit octonions with $N^2 >0$ having time-like vector part:
\begin{equation} \label{R(J)}
e^{J_n \theta_n } = \cosh \theta_n + J_n \sinh \theta_n ~,
\end{equation}
where $\theta_n$ are three real hyperbolic angles. Let us consider the example when $\epsilon = J_1$ in (\ref{Rs}). The equivalent representations of $J_1$ in the algebra (\ref{algebra}) are:
\begin{equation} \label{J1}
J_1 = J_2 j_3 = - J_3j_2 = j_1I~,
\end{equation}
or
\begin{equation} \label{J_1}
J_1J_2 = -j_3 ~~~~~ J_1j_2 = J_3~, ~~~~~ J_1I = j_1~,
\end{equation}
Then (\ref{s}) can be rewritten as:
\begin{equation} \label{s-J1}
s = \sqrt{\omega^2 - \lambda_1^2} e^{J_1 \theta_\omega} + \sqrt{x_3^2 - \lambda_2^2} e^{-J_1 \theta_\lambda}J_2 + \sqrt{x_2^2 - \lambda_3^2} e^{J_1 \theta_x}j_2 + \sqrt{x_1^2 - t^2} e^{J_1 \theta_t}I ~,
\end{equation}
where the angles are given by:
\begin{eqnarray}
&\cosh \theta_\omega = \frac {\omega }{\sqrt{\omega^2 - x_1^2}}~, ~~~~~ \cosh \theta_\lambda = \frac {\lambda_2}{\sqrt{x_3^2 - \lambda_2^2}}~, \nonumber \\
&\cosh \theta_x = \frac {x_2 }{\sqrt{x_2^2 - \lambda_3^2}}~, ~~~~~ \cosh \alpha_t = \frac {t }{\sqrt{x_1^2 - t^2}} ~.
\end{eqnarray}
So when $R = e^{J_1\theta_1}$, the product (\ref{Rs}) corresponds to the simultaneous rotations in the planes $(\omega, \lambda_1)$, $(\lambda_2, x_3)$, $(x_2, \lambda_3)$ and $(x_1, t)$ by the hyperbolic angle $\theta_1$. Similar results we have for the rotations by the other two vector-like units, $J_2$ and $J_3$.}

\item {The last vector-like basis element, $I$, also has the positive square, $I^2 = 1$, and to represent corresponding transformations we need the unit octonion with $N^2 > 0$ having time-like vector part:
\begin{equation} \label{R(I)}
e^{I \sigma } = \cosh \sigma + I \sinh \sigma ~,
\end{equation}
where $\sigma$ is the real hyperbolic angle. When in (\ref{Rs}) we put $\epsilon = I$, we can use the expressions (\ref{I}) of the equivalent representations of $I$, or
\begin{equation} \label{I_}
Ij_1 = - J_1 ~, ~~~~~ Ij_2 = - J_2 ~, ~~~~~ Ij_3 = - J_3 ~.
\end{equation}
Then the split octonion (\ref{s}) can be rewritten as:
\begin{equation} \label{s-I}
s = \sqrt{\omega^2 - t^2} e^{I \sigma_\omega} + \sqrt{x_1^2 - \lambda_1^2} e^{-I \sigma_1}j_1 +\sqrt{x_2^2 - \lambda_2^2} e^{-I \sigma_2}j_2 + \sqrt{x_3^2 - \lambda_3^2} e^{-I \sigma_3}j_3 ~,
\end{equation}
where four hyperbolic angles are given by:
\begin{eqnarray}
&\cosh \sigma_\omega = \frac {\omega}{\sqrt{\omega^2 - t^2}}~, ~~~~~ \cosh \sigma_1 = \frac {x_1}{\sqrt{x_1^2 - \lambda_1^2}} ~, \nonumber \\
&\cosh \sigma_2 = \frac {x_2}{\sqrt{x_2^2 - \lambda_2^2}} ~, ~~~~~ \cosh  \sigma_3 = \frac {x_3}{\sqrt{x_3^2 - \lambda_3^2}} ~.
\end{eqnarray}
So the left product (\ref{Rs}), when $R = e^{I\sigma}$, corresponds to the simultaneous rotations in the planes $(\omega, t)$, $(x_1, \lambda_1)$, $(x_2, \lambda_2)$ and $(x_3, \lambda_3)$ by the hyperbolic angle $\sigma$.}
\end{itemize}

So the one-side products (\ref{Rs}) of a octonion $s$ by each of the seven operators (\ref{R(j)}), (\ref{R(J)}) and (\ref{R(I)}) yield simultaneous rotations in four mutually orthogonal planes of the octonionic (4+4)-space by same angles \cite{Man-Sch}. One of these four planes is formed by the unit element, $1$, and the hypercomplex basis unit that defines the axis of rotation. The remaining orthogonal planes are given by three pairs of other basis elements which form associative triplets with the chosen basis unit.

The decomposition of a split octonion in the form (\ref{s-j1}) is valid only if its norm (\ref{sN}) is positive. In contrast with uniform rotations around $j_n$, we have limited rotations, (\ref{s-J1}) and (\ref{s-I}), around $J_n$ and $I$. For these cases we should require also positiveness of the 2-norms of the each of the four planes of rotations. For example, for (\ref{s-I}) the expressions of 2-norms of four orthogonal planes are: $\sqrt{\omega^2 - t^2}, \sqrt{x_1^2 - \lambda_1^2}, \sqrt{x_2^2 - \lambda_2^2}$ and $\sqrt{x_3^2 - \lambda_3^2}$. These hyperbolic properties are the result of the existence of zero divisors in split algebras (\ref{Zero}).


\section{Automorphisms and G2} \label{Automorphisms}
\setcounter{equation}{0}

From the expressions of a hypercomplex unit of split octonions by two other basis elements, which form associative triplets with the selected unit, one can find orthogonal to it planes of rotations. An automorphism (\ref{U}) for each basis unit introduces two angles of rotation in these planes. So there are seven independent automorphisms each by two angles that correspond to $2 \times 7 = 14$ generators of the algebra $G_2^{NC}$. For example, the automorphism by the two real compact angles, $\alpha_1$ and $\beta_1$, which leaves unchanged the basis unit $j_1 = J_2J_3 = j_2j_3 = J_1I$, has the form:
\begin{eqnarray}
j_1' &=& j_1~, \nonumber \\
j_2' &=& j_2\cos \alpha_1 + j_3\sin \alpha_1 ~, \nonumber\\
j_3' &=& j_1'j_2'= j_3\cos \alpha_1 - j_2\sin \alpha_1 ~, \nonumber \\
J_1' &=& J_1\cos \beta_1 + I\sin \beta_1  ~, \label{rotation} \\
I' &=& J_1'j_1' = I \cos \beta_1 - J_1 \sin \beta_1 ~, \nonumber \\
J_2' &=& j_2'I' = J_2\cos (\alpha_1 - \beta_1) + J_3\sin (\alpha_1 - \beta_1) ~, \nonumber \\
J_3' &=& j_3'I' = J_3\cos (\alpha_1 - \beta_1) - J_2\sin (\alpha_1 - \beta_1) ~. \nonumber
\end{eqnarray}
It is known that automorphisms in the octonionic algebra are completely defined by the images of three basis elements that do not form associative subalgebras ($j_1$, $j_2$ and $J_1$ in the case of (\ref{rotation})). By the definition (\ref{U}) an automorphism does not affect the scalar part of $s$, while the images of the other hypercomplex elements ($j_3$, $I$, $J_2$ and $J_3$ in (\ref{rotation})) are determined using the algebra (\ref{algebra}). Then it is obvious that the transformed basis $J_n', j_n', I'$ satisfy the same multiplication rules as $J_n, j_n, I$.

Similar to (\ref{rotation}) there exists two automorphisms with fixed $j_2 = J_3J_1 = j_3j_1 = J_2I$ and $j_3 = J_1J_2 = j_1j_2 = J_3I$ axes, each generating the two compact angles, $\alpha_2, \beta_2$ and $\alpha_3, \beta_3$. Three different representations of a basis unit indicate the planes of corresponding automorphisms and the positive directions of rotations.

One can define also hyperbolic rotations (automorphisms) around the three vector-like units $J_n$ by the angles $\theta_n$ and $\phi_n$. For example, similar to the transformation (\ref{rotation}) automorphism around the axis $J_1 = j_3J_2 = J_3j_2 = j_1I $ is:
\begin{eqnarray}
J_1' &=& J_1~, \nonumber \\
J_2' &=& J_2 \cosh (\phi_1 + \theta_1)/2 + j_3 \sinh (\phi_1 + \theta_1)/2 ~, \nonumber\\
j_3' &=& J_1'J_2' = j_3\cosh (\theta_1 + \phi_1)/2 + J_2\sinh (\theta_1 + \phi_1)/2 ~, \nonumber\\
I' &=& I \cosh \theta_1 - j_1 \sinh \theta_1 ~, \label{boost} \\
j_1' &=& J_1'I' = j_1 \cosh \theta_1 - I \sinh \theta_1 ~, \nonumber \\
j_2' &=& J_2'I' = j_2\cosh (\phi_1 - \theta_1)/2 + J_3 \sinh (\phi_1 - \theta_1)/2 ~, \nonumber \\
J_3' &=& j_3'I' = J_3\cosh (\phi_1 - \theta_1)/2 + j_2\sinh (\phi_1 - \theta_1)/2 ~, \nonumber
\end{eqnarray}
were we chose the representation with half-angle rotations in order to write the transformations of $\lambda_n$ and $x_n$ in symmetric form. The automorphisms with the fixed $J_2 = j_1J_3 = J_1j_3 = j_2I$ and $J_3 = j_2J_1 = J_2j_1 = j_3I$ axes generate hyperbolic angles, $\theta_2, \phi_2$ and $\theta_3, \phi_3$, respectively.

Finally for the rotations around the time direction, $I = J_1j_1 = J_2j_2 = J_3j_3 $, we have only two hyperbolic angles, $k_1$ and $k_2$:
\begin{eqnarray}
I' &=& I~, \nonumber \\
j_1' &=& j_1 \cosh k_1 + J_1 \sinh k_1 ~, \nonumber\\
j_2' &=& j_2 \cosh k_2 + J_2 \sinh k_2 ~, \nonumber \\
J_1' &=& j_1'I' = J_1\cosh k_1 + j_1\sinh k_1 ~, \label{uncertainty}\\
J_2' &=& j_2'I' = J_2 \cosh k_2 + j_2 \sinh k_2 ~, \nonumber \\
j_3' &=& j_1'j_2' = j_3 \cosh (k_1 + k_2) - J_3 \sinh (k_1 + k_2) ~, \nonumber \\
J_3' &=& j_3'I' = J_3 \cosh (k_1 + k_2) - j_3\sinh (k_1 + k_2) ~. \nonumber
\end{eqnarray}
In order to present the transformations of $\lambda_n$ and $x_n$ in (\ref{x-nu}) in the symmetric form, it is convenient to introduced instead of $k_1$ and $k_2$ the three hyperbolic angles $\varphi_n$,
\begin{equation} \label{k-varphi}
k_1 = \varphi_1 - \frac 13 \sum_n \varphi_n~, ~~~~~ k_2 = \varphi_2 - \frac 13 \sum_n \varphi_n~, ~~~~~ -(k_1 + k_2) = \varphi_3 - \frac 13 \sum_n \varphi_n~.
\end{equation}


\section{Generators of G2} \label{Generators}
\setcounter{equation}{0}

The group $G_2^{NC}$ first was studied by E. Cartan in his thesis \cite{Cart}. He considered (3+4)-space with the seven coordinates, $(y^n, t, z^n)$, and linear operators of their transformations,
\begin{eqnarray} \label{X}
&X_{nn} = - z_n \frac {\partial}{\partial z_n} + y_n \frac{\partial}{\partial y_n} + \frac 13 \sum^3_{m=1}\left( z_m\frac {\partial}{\partial z_m} - y_m \frac{\partial}{\partial y_m} \right), ~~~~~ ({\rm no~ summing~ over}~ n) \nonumber\\
&X_{n0} = - 2t \frac{\partial}{\partial z_n} + y_n \frac{\partial}{\partial t} + \frac 12 \varepsilon_{nmk} \left( z^m\frac {\partial}{\partial y_k} - z^k \frac{\partial}{\partial y_m}\right) , \nonumber\\
&X_{0n} = - 2t \frac {\partial}{\partial y_n} + z_n \frac {\partial}{\partial t} + \frac 12 \varepsilon_{nmk} \left( y^m \frac{\partial}{\partial z_k} - y^k \frac {\partial}{\partial z_m}\right) ,\\
&X_{nm} = - z_m\frac {\partial}{\partial z_n} + y_n \frac {\partial}{\partial y_m}, ~~~~~ (n \neq m) \nonumber
\end{eqnarray}
which preserves the quadratic form $dt^2 + z_ny^n$. Totally there are 15 generators in (\ref{X}) since $n,m = 1,2,3$, but the operators $X_{nn}$ are linearly dependent,
\begin{equation}
X_{11} + X_{22} + X_{33} = 0 ~,
\end{equation}
which explains why $G_2^{NC}$ is 14- and not 15-dimensional.

By transition to our coordinates, $x_n$ and $\lambda_n$,
\begin{eqnarray} \label{lambda,x}
&y_n = \lambda_n + x_n~, ~~~~~ \frac {\partial}{\partial y_n} = \frac 12 \left(\frac {\partial}{\partial \lambda_n} + \frac {\partial}{\partial x_n}\right)~,\nonumber \\
&z_n = \lambda_n - x_n~, ~~~~~ \frac {\partial}{\partial z_n} = \frac 12 \left(\frac {\partial}{\partial \lambda_n} - \frac {\partial}{\partial x_n}\right)~,
\end{eqnarray}
from (\ref{X}) we can find the operators of the transformation of the vector part $V$ of (\ref{s}) which preserves the diagonal quadratic form, $V^2 = \lambda_n\lambda^n + dt^2 - x_nx^n$:
\begin{eqnarray}
&X_{nn} = x_n \frac {\partial}{\partial x_n} + \lambda_n \frac{\partial}{\partial \lambda_n} - \frac 13 \sum^3_{m=1}\left( x_m\frac {\partial}{\partial x_m} + \lambda_m \frac{\partial}{\partial \lambda_m} \right),  \nonumber\\
&X_{n0} = \left(\lambda_n \frac{\partial}{\partial t} - t \frac {\partial}{\partial \lambda_n}\right) + \left(x_n \frac {\partial}{\partial t} + t \frac {\partial}{\partial x_n}\right) + \nonumber \\
&+ \frac 14 \varepsilon_{nmk} \left( \lambda^m \frac {\partial}{\partial \lambda_k} - \lambda^k \frac{\partial}{\partial \lambda_m}\right) - \frac 14 \varepsilon_{nmk} \left( x^m \frac {\partial}{\partial x_k} - x^k \frac{\partial}{\partial x_m}\right) + \nonumber \\
&+ \frac 14 \varepsilon_{nmk} \left( \lambda^m \frac {\partial}{\partial x_k} + x^k \frac{\partial}{\partial \lambda_m}\right) -
\frac 14 \varepsilon_{nmk} \left( x^m \frac {\partial}{\partial \lambda_k} + \lambda^k \frac{\partial}{\partial x_m}\right),  \nonumber\\
&X_{0n} = \left(\lambda_n \frac{\partial}{\partial t} - t \frac {\partial}{\partial \lambda_n}\right) - \left(x_n \frac {\partial}{\partial t} + t \frac {\partial}{\partial x_n}\right) + \label{X-nm} \\
&+ \frac 14 \varepsilon_{nmk} \left( \lambda^m \frac {\partial}{\partial \lambda_k} - \lambda^k \frac{\partial}{\partial \lambda_m}\right) - \frac 14 \varepsilon_{nmk} \left( x^m \frac {\partial}{\partial x_k} - x^k \frac{\partial}{\partial x_m}\right) - \nonumber \\
&- \frac 14 \varepsilon_{nmk} \left( \lambda^m \frac {\partial}{\partial x_k} + x^k \frac{\partial}{\partial \lambda_m}\right) +
\frac 14 \varepsilon_{nmk} \left( x^m \frac {\partial}{\partial \lambda_k} + \lambda^k \frac{\partial}{\partial x_m}\right), \nonumber \\
&X_{nm} = - \frac 12 \left( \lambda_n \frac {\partial}{\partial \lambda_m} - \lambda_m \frac {\partial}{\partial \lambda_n}\right) - \frac 12 \left( x_n \frac {\partial}{\partial x_m} - x_m \frac {\partial}{\partial x_n}\right) + \nonumber \\
&+ \frac 12 \left( x_n \frac {\partial}{\partial \lambda_m} + \lambda_m \frac {\partial}{\partial x_n}\right) +
\frac 12 \left( \lambda_n \frac {\partial}{\partial x_m} + x_m \frac {\partial}{\partial \lambda_n}\right). ~~~~~ (n \neq m) \nonumber
\end{eqnarray}


\section{Zero divisors} \label{Zero}
\setcounter{equation}{0}

In the algebra of split octonions two types of zero divisors, idempotent elements (projection operators) and nilpotent elements (Grassmann numbers), can be constructed \cite{Sc, So}.

There exist four non-commuting (totally eight) primitive idempotents,
\begin{equation} \label{D}
D^{\pm (J)}_n = \frac 12 \left(1 \pm J_n \right)~,~~~~~~~ D^{\pm (I)} = \frac 12 \left(1 \pm I\right) ~, ~~~~~ (n = 1,2,3)
\end{equation}
which obey the relations:
\begin{eqnarray} \label{DD}
D^{\pm (J)}_n D^{\pm (J)}_n = D^{\pm (J)}_n~, ~~~~~ D^{+(J)}_nD^{- (J)}_n = 0 ~, \nonumber \\
D^{\pm (I)}D^{\pm (I)} = D^{\pm (I)}~, ~~~~~D^{+(I)}D^{- (I)} = 0~,
\end{eqnarray}

We have also the four non-commuting classes (totally twelve) of primitive nilpotents,
\begin{equation} \label{G}
G^{\pm (J)}_n = \frac 12 \left(J_n \pm j_n\right) ~, ~~~~~ G^{\pm (I)}_n = \frac 12 \left( I \pm j_n \right) ~, ~~~~~ (n = 1,2,3)
\end{equation}
with the properties:
\begin{eqnarray} \label{GG}
G^{\pm (J)}_n G^{\pm (J)}_n = 0~, ~~~~~ G^{\pm (J)}_n G^{\mp (J)}_n = D^{\mp (I)} ~, \nonumber \\
G^{\pm (I)}_n G^{\pm (I)}_n = 0~, ~~~~~ G^{\pm (I)}_n G^{\mp (I)}_n = D^{\pm (J)}_n ~.
\end{eqnarray}
From this relations we see that separately nilpotents are Grassmann numbers, but different $G_n$-s do not commute with each other, in contrast to the projection operators (\ref{DD}). Instead the quantities $G^\pm_n$ are elements of so-called algebra of Fermi operators with the anti-commutators:
\begin{eqnarray}
G^{\pm (J)}G^{\mp (J)} + G^{\mp (J)}G^{\pm (J)} = 1~, \nonumber \\
G^{\pm (I)}G^{\mp (I)} + G^{\mp (I)}G^{\pm (I)} = 1~.
\end{eqnarray}
The algebra of Fermi operators is some syntheses of the Grassmann and Clifford algebras.

For the completeness we note that the idempotents and nilpotents obey the following algebra:
\begin{eqnarray}
D^{\pm (J)}_nG^{\pm (I)}_n = G^{\pm (I)}_n~, ~~~~~ D^{\pm (J)}_nG^{\mp (I)}_n = 0~, \nonumber \\
D^{\pm (I)}G^{\pm (J)}_n = 0~, ~~~~~ D^{\pm (I)}G^{\mp (J)}_n = G^{\mp (J)}_n~.
\end{eqnarray}

Using commuting zero divisors any octonion (\ref{s}) can be written in the two equivalent forms:
\begin{eqnarray} \label{sDec}
s &=& D^{+(J)}_n \left(\frac 13 \omega + \lambda^n \right) + G^{+(I)}_n \left(ct + x^n \right) + D^{-(J)}_n \left(\frac 13 \omega - \lambda^n \right) + G^{-(I)}_n\left(ct - x^n  \right) ~, \nonumber \\
s &=& D^{+(I)} \left(\omega + ct\right) + G^{+(J)}_n \left(\lambda^n + x^n \right) + D^{-(I)} \left(\omega - ct\right) + G^{-(J)}_n \left(\lambda^n - x^n \right) ~.
\end{eqnarray}
The norm of a split octonion (\ref{V2}) contains two types of 'light-cones' and we can introduce two types of critical signals (elementary particles) visualized in the decompositions (\ref{sDec}).


\end{document}